\documentclass{article}

\usepackage{amssymb}                                    
\usepackage{mathrsfs}                    
\usepackage{amsmath}                    
\usepackage{amsfonts}                   
\usepackage{amsfonts}                   
\usepackage{amsthm}                     
\usepackage{hyperref}

\usepackage{color}

\usepackage{fullpage}

\usepackage[all]{xy}

\makeindex

\theoremstyle{plain}

\theoremstyle{definition}

\numberwithin{equation}{section}

\renewcommand{\(}{\begin{equation}}
\renewcommand{\)}{\end{equation}}
\newcommand{\bea}{\begin{eqnarray}}
\newcommand{\eea}{\end{eqnarray}}

\def\proofoftheorem #1 {{Proof of theorem \ref{#1}.}\hspace{7pt}}

\begin{document}

\title{The WZW term of the M5-brane and differential cohomotopy}
\author{D. Fiorenza, H. Sati and U. Schreiber}
\maketitle

\begin{abstract}
We combine rational homotopy theory and higher Lie theory to describe the
 WZW term in the M5-brane sigma model.
  We observe that
 this term admits a natural interpretation as a twisted 7-cocycle on super-Minkowski spacetime with coefficients in the rational 4-sphere. This exhibits the WZW term as an element in twisted cohomology, with the twist given by the cocycle of the M2-brane.
  We consider integration of this rational situation to differential cohomology
  and differential cohomotopy.
%

\end{abstract}

\tableofcontents

\section{Introduction and survey}

The Green-Schwarz-type sigma-models for super $p$-branes with manifest $N$-supersymmetry on
$d$-dimensional target space are higher-dimensional and supergeometric WZW-models \cite{HenMez}.
They are induced by the non-trivial Lorentz-invariant $(p+2)$-cocycles $\mu_{p+2}$ on super Minkowski
spacetime $\mathbb{R}^{d-1,1\vert \mathbf{N}}$, regarded as the supertranslation super Lie algebra,
with coefficients in the trivial 1-dimensional module. The left-invariant extension of these cocycles to closed differential forms
are locally the differentials  of the WZW-term Lagrangian in the sigma-model. Globally, they are the curvature forms of the 
WZW $p$-gerbe connections.

\medskip
Or rather, this is the case for the $p$-branes in the ``old brane scan'' \cite{AT}
which are those on which no other branes may end. Equivalently these are those branes without (higher) gauge fields
on their worldvolume, e.g. the fundamental string and the membrane,
but not the D-branes and not the M5-brane. The latter are instead defined by
analogous algebraic structures but on ``extended super Minkowski spacetimes'' \cite{CdAIP}.
In \cite{InfinityWZW} 
these were identified with  super-$L_\infty$-algebras  cocycles defined not on the super-Lie algebra of translations on the super-Minkowski space time, but rather on some
super-$L_\infty$-extension of it, induced by one ``lower level'' cocycle. 

\medskip
Given a pair of such consecutive super-$L_\infty$-cocycles, with one defined on the homotopy
fiber of the former, then a natural question is whether the second cocycle carries a suitable equivariance (up to coherent higher homotopy)
making it descent back to ordinary (i.e., non-extended) super-Minkowski spacetime. When this happens, however, the descended cocycle will be
a cocycle with nontrivial coefficients, defining a class in the twisted cohomology of super-Minkowski spacetime (with the twist given by the first cocycle), see 
\cite[section 4]{NSSa}.
For the case of the super D-$p$-brane cocycles defined on the extension of super-Minkowski spacetime
induced by the superstring cocycle this is a familiar situation: the D-brane charges are indeed in a
cohomology theory, namely K-theory, which is twisted by the cohomology class of the string's background gauge field, the
B-field. The point to note here is that at the rational level, this famous insight may already be
deduced from just the higher super-Lie theory of super-Minkowski spacetime.

\medskip
In view of this, here we analyze the analogous situation for the M5-brane WZW term, as considered in \cite{BLNPST1},
defined on the extension of 11-dimensional super-Minkowski spacetime induced by the M2-brane cocycle.
In this context the proper cohomological nature of the charges, which is the M-theoretic analog of the lift from ordinary
cohomology to K-theory for D-brane charges,
remains a not completely solved question;  see however the proposals in \cite{KS1}
\cite{Sa}. We find that, indeed, the cocycle defining the M5-brane WZW term carries equivariant
structure that makes it descend back to unextended super-Minkowski spacetime where
it takes coefficients in (a differential refinement of) the rational 4-sphere. The topological counterpart to this phenomenon has been
observed in \cite{Sati13}. Via the Hopf fibration this exhibits a twisted cohomology
theory with coefficients in degree 7, twisted by cohomology in degree 4.

\medskip
In going from this local situation to a global situation where a curved super-spacetime is only locally isomorphic to super-Minkowski spacetime, one passes from global differential form data for the fields to local satisfying suitable coherence conditions on the overlaps of the local charts. Mathematically, this amounts to considering a (higher) stack of fields which is not a bare sheaf, as for plain differential forms, but that comes with the data of nontrivial and higher gauge transformations. In this global picture the WZW cocycles become elements in twisted differential nonabelian cohomology, with coefficients in a suitable stack of fields.
Due to the above appearance of the rational Hopf fibration, there is a natural candidate for such a stack, given by a differential refinement $S^4_{\mathrm{conn}}$ of the homotopy type of the 4-sphere.
%
%

\medskip
Here the appearance of the unstabilized 4-sphere instead of, say, the 4-suspended sphere spectrum is crucial,
since it is the former that has a non-torsion contribution not only in degree-4 (for the supergravity C-field)
but also in degree 7 (for its electromagnetic/Hodge dual).
Since nonabelian cohomology with coefficients in unstabilized spheres is also known as
\emph{cohomotopy}, one may regard this as realizing the supergravity higher gauge fields as cocycles
in ``differential cohomotopy''. This has been suggested before in \cite[section 2.5]{Sati13}.
We close with an outlook on the definite globalization of these cocycles over curved super-spacetimes.
More details on this may be found in \cite{WZWterms}, with a full discussion to appear elsewhere.

\section{The M5-brane cocycle}

Let $d\geq 1$ be a positive integer and let $N$ be a real representation of the group  $\mathrm{Spin}(d-1,1)$.
We write $\mathbb{R}^{d-1,1\vert N}$ for the corresponding super-Minkowski spacetime.
Also, we denote by  $\overline{(-)}\Gamma (-) : N \otimes N \to \mathbb{R}^{d}$  the symmetric
bilinear $\mathrm{Spin}$-equivariant
pairing associated with the real representation $N$. See \cite{InfinityWZW} for further pointers.

\medskip
Following the
conventions of \cite{DF}, we write $\{e^a\}_{a = 1}^{d}$, $\{\psi^\alpha\}_{\alpha = 1}^{\mathrm{dim}_\mathbb{R}(N)}$ for
a basis of left-invariant 1-forms on $\mathbb{R}^{d-1,1\vert N}$, with respect to the super-translation action.
The 1-forms $e^a$'s and $\psi^\alpha$'s are in bidegrees  $(1,\mathrm{even})$ and
$(1,\mathrm{odd})$, respectively. As an instance of non-abelianity of super-translations, left-invariant 1-forms on super-Minkowski space are not necessarily closed; rather they satisfy
$$
  d \psi^\alpha = 0
  \;\;
  \,,
  \;\;
  d e^a = \overline{\psi}\wedge \Gamma^a \psi
  \,,
$$
where $\psi$ is the $N$-valued 1-form whose components with respect to a fixed linear basis of $N$ are the $\psi^\alpha$.
The \emph{Chevalley-Eilenberg algebra} of the super-Minkowski spacetime $\mathbb{R}^{d-1,1\vert N}$ is then defined as the differential $(\mathbb{N},\mathbb{Z}_2)$-bigraded commutative algebra of left-invariant polynomial differential forms on $\mathbb{R}^{d-1,1\vert N}$. Algebraically, it is the polynomial algebra (i.e., the free bigraded commutative algebra) $\mathbb{R}[e^a,\psi^\alpha]$ on the generators $\{e^a,\psi^\alpha\}$ in bidegrees $(1,\mathrm{even})$
and $(1,\mathrm{odd})$, respectively, endowed with the differential $d$ which is defined as above on the generators and then extended to all polynomials by the graded Leibniz rule, with $d$ of bidegree $(1,\mathrm{even})$.
We will denote this Chevalley-Eilenberg algebra by the symbol
$\mathrm{CE}(\mathbb{R}^{d-1,1\vert N})$. Notice that $\mathrm{CE}(\mathbb{R}^{d-1,1\vert N})$ is free as a bigraded commutative algebra, but not as a differential bigraded commutative algebra. One refers to this fact by saying that $\mathrm{CE}(\mathbb{R}^{d-1,1\vert N})$ is a \emph{semifree} differential (bi-)graded commutative algebra; with some abuse of terminology, these algebras are also known as
``free differential algebras'' (FDAs) in parts of the physics literature.

\medskip
In terms of the (super-)$L_\infty$-algebras/semifree differential (bi-)graded commutative algebras duality
(see for instance \cite{SSSI} for review well-adapted to the present discussion), the algebra $\mathrm{CE}(\mathbb{R}^{d-1,1\vert N})$ precisely encodes the super-Lie algebra structure of the left-translation in the super-Minkowski spacetime $\mathbb{R}^{d-1,1\vert N}$. As a matter of notation, we write $\mathrm{CE}(\mathfrak{g})$ for the semifree differential (bi-)graded commutative algebra dual to a given (super-)$L_\infty$-algebra $\mathfrak{g}$. Notice that $\mathrm{CE}$ is a contravariant equivalence between the category of (super-)$L_\infty$-algebras and that of semifree differential (bi-)graded commutative algebras. In particular, an $L_\infty$-morphism $f\colon \mathfrak{g}\to \mathfrak{h}$ is equivalently a morphism of differential (bi-)graded commutative algebras $f^*\colon\mathrm{CE}(\mathfrak{h})\to \mathrm{CE}(\mathfrak{g})$.  Moreover, the contravariant equivalence $\mathrm{CE}$ allows one to translate homotopical constructions on the $L_\infty$-algebra side into homotopical constructions on the differential graded commutative algebra side, and vice versa.
See, e.g.,  \cite{LocalObservables} for details.

\medskip
We now go back to our explicit cocycles. For any nonnegative integer $p$, write
$$
  \mu_{p+2} := \overline{\psi} \wedge \Gamma^{a_1 \cdots a_p} \psi\wedge e_{a_1} \wedge \cdots \wedge e_{a_p}
 $$
for the canonical Lorentz-invariant degree-$(p+2)$ element in $\mathrm{CE}(\mathbb{R}^{d-1,1\vert N})$. Notice that $\mu_{p+2}$ has even spinor-degree and bosonic degree $p$. That is, when $\mu_{p+2}$ is closed, i.e., $d\mu_{p+2}=0$, it can be interpreted as a morphism of differential (bi-)graded commutative algebras $\mathbb{R}[g_{p+2}]\to \mathrm{CE}(\mathbb{R}^{d-1,1\vert N})$, where $\mathbb{R}[g_{p+2}]$ is the polynomial algebra on a single generator $g_{p+2}$ of degree $(p+2,\mathrm{even})$, with the zero differential. Viewed in the dual (super-)$L_\infty$-algebra picture, $\mathbb{R}[g_{p+2}]$ corresponds to the graded vector space $\mathbb{R}[p+1]$ consisting of the vector space $\mathbb{R}$ placed in degree $p+1$, and endowed with the trivial $L_\infty$-brackets $[\dots]_k=0$ for every $k$. Therefore, a closed $\mu_{p+2}$ is equivalently a super-Lie algebra $(p+1)$-cocycle
\[
\mu_{p+2}\colon \mathbb{R}^{d-1,1\vert N}\to \mathbb{R}[p+1].
\]
If $\mu_{p+2}$ is a cocycle, then it defines a cohomology class in the cohomology of the Chevalley-Eilenberg algebra $\mathrm{CE}(\mathbb{R}^{d-1,1\vert N})$.
The triples $(d,N,p)$ for which these classes are nontrivial constitute
the entries of the \emph{old brane scan} \cite{AT}. For example for $d =11$, $N = \mathbf{32}$ and
$p = 2$ one has that $\mu_4$ in $\mathrm{CE}(\mathbb{R}^{10,1\vert \mathbf{32}})$ is such an element
and it gives the curvature of the WZW term for the Green-Schwarz type sigma-model of the
M2-brane \cite{BST}. The situation for the M5-brane is more interesting:
the two elements
$\mu_4, \mu_7 $ in $ \mathrm{CE}(\mathbb{R}^{10,1\vert \mathbf{32}})$
satisfy {\cite[(3.28)]{DF}}
$$
  d \mu_4 = 0
  \;\;\,,
  \;\;
  d \mu_7 = 15 \, \mu_4 \wedge \mu_4
  \,.
$$

\medskip
Hence, while $\mu_7$ is not itself closed, a closed element may be obtained from it
if we adjoin a generator $h_3$ in degree 3 and declare that $d h_3 = - 15\,\mu_4$.
The ``space'' defined as having such an $h_3$-extended algebra of left invariant forms has been called
``extended super-Minkowski spacetime'' \cite{CdAIP}. In terms of the homotopy theory of
super-$L_\infty$-algebras this construction of an extended super-Minkowski space corresponds to considering the
 super-$L_\infty$-algebra extension induced  by the cocycle $\mu_{4}$, i.e., the homotopy fiber of $\mu_4\colon \mathrm{CE}(\mathbb{R}^{10,1\vert \mathbf{32}})\to \mathbb{R}[3]$, see \cite[theorem 3.1.13]{LocalObservables}. Following \cite{InfinityWZW} we write $\mathfrak{m}2\mathfrak{brane}$ for the super Lie 3-algebra which dually corresponds to
having added the generator $h_3$ with $d h_3 = - \mu_4$ to the Chevalley-Eilenberg algebra of $\mathbb{R}^{10,1\vert \mathbf{32}}$:
$$
  \mathrm{CE}(\mathfrak{m}2\mathfrak{brane}) = (\mathrm{CE}(\mathbb{R}^{10,1\vert \mathbf{32}})\otimes \mathbb{R}[h_3], d h_3 = - \mu_4)
  \,.
$$
Then the degree 7 element
$$
  h_3 \wedge \mu_4 + \tfrac{1}{15} \mu_7 \;\;
$$
in $\mathrm{CE}(\mathfrak{m}2\mathfrak{brane})$ is closed.
%
%
%
Translating this back in the language of super-$L_\infty$-algebras we obtain 
a homotopy pullback of super-$L_\infty$-algebras
of the form
\begin{equation}\label{The7CocycleOnHomotopyFiber}
  \raisebox{20pt}{
  \xymatrix{
    \mathfrak{m}2\mathfrak{brane} \ar[d]\ar[rr]^-{h_3 \wedge \mu_4 + \tfrac{1}{15} \mu_7} && \mathbb{R}[6]\ar[rr]&& 0\ar[d]
    \\
    \mathbb{R}^{10,1\vert \mathbf{32}}
    \ar[rrrr]^{\mu_4} && && \mathbb{R}[3]\;.
  }}
\end{equation}

The degree-7 element $h_3 \wedge \mu_4 +  \tfrac{1}{15}\mu_7$ is the curvature of the WZW term for the M5-brane propagating
in a flat spacetime with vanishing background $C$-field, according to \cite[(9)]{BLNPST1}.
Notice that this is a higher-degree analog of the curvature 3-form
$F^a \wedge \lambda_a + \langle \theta \wedge [\theta \wedge \theta]\rangle$ for the standard gauged 2d WZW model
\cite[(A.14)]{Witten92},
with the degree 3 generator $h_3$ being the higher analog of the degree 2 generators $F^a$ in the Cartan model for equivariant de Rham cohomology.
In this sense the M5-brane WZW model is a \emph{higher gauged} WZW model, carrying not a 1-form but a 2-form gauge field
on its worldvolume. 

\medskip
While on (extended) super-Minkowski
spacetime these WZW Lagrangian are given by globally defined differential forms, if we move to a curved super-spacetime $X$ then we only have local differential form WZW Lagrangians. These do not necessarily glue together into a single differential form, but still have a coherent compatibility on local charts overlaps for an atlas exhibiting $X$ as locally diffeomorphic to (extended) super-Minkowski spacetime: they are a cocycle in the Deligne cohomology of $X$. From a geometric point of view, they define a \emph{higher $U(1)$-bundle with connection} on the (extended) super-spacetime $X$.  Following \cite{FSS}, we write
$\mathbf{B}^{p+1}U(1)_{\mathrm{conn}}$ for the $(p+1)$-stack of $(p+1)$-$U(1)$-bundles with connection. This $(p+1)$-stack comes equipped with a natural ``forget the connection'' morphism $\mathbf{B}^{p+1}U(1)_{\mathrm{conn}}\to \mathbf{B}^{p+1}U(1)$ to the $(p+1)$-stack of principal $U(1)$-$(p+1)$-bundles (or $p$-bundle gerbes), which is in turn a smooth refinement of the Eilenberg-MacLane space $K(\mathbb{Z},p+2)$. 
With this in mind, 
the homotopy commutative diagram
(\ref{The7CocycleOnHomotopyFiber}) translates to the following geometric picture for WZW Lagrangians
(details are in \cite{WZWterms}): 
%
%
%
%
let $X$ be a super-spacetime locally modeled on $\mathbb{R}^{10,1\vert \mathbf{32}}$ and let $\mathbf{L}_{\mathrm{WZW}}^{M2}\colon X \to \mathbf{B}^{3}U(1)_{\mathrm{conn}}$
be the globalized WZW Lagrangian for the M2-brane. Denote by $\tilde X$ the 3-form twisted 
homotopy fiber of $\mathbf{L}_{\mathrm{WZW}}^{M2}$.
Then $\tilde{X}$ comes with a natural projection to the total space $\hat{X}$ of the 2-gerbe underlying the M2-term, and we have a natural homotopy commutative diagram of the form
\begin{equation}
  \raisebox{20pt}{
  \xymatrix{
   \tilde X \ar[d] \ar[rr]^-{{\mathbf{L}_{\mathrm{WZW}}^{\mathrm{M5}}}} 
   && 
   \mathbf{B}^6U(1)_{\mathrm{conn}} \ar[rr] && 0 \ar[d]
    \\
    X
    \ar[rr]^-{{\mathbf{L}_{\mathrm{WZW}}^{\mathrm{M2}}}} 
    && \mathbf{B}^3U(1)_{\mathrm{conn}}
    \ar[rr]
    &&
    \mathbf{B}(\mathbf{B}^2 U(1)_{\mathrm{conn}})
    \;,
  }}
  \label{HigherGaugedTargetSpacetime}
\end{equation}
where ${\mathbf{L}_{\mathrm{WZW}}^{\mathrm{M5}}}$ is the globalized WZW Lagrangian for the M5-brane, and where $\mathbf{B}^3U(1)_{\mathrm{conn}}
   \to \mathbf{B}(\mathbf{B}^2 U(1)_{\mathrm{conn}})$ is the morphism that forgets the 3-form part of a 3-connection; see  \cite{FiorenzaSatiSchreiberCS}.
%
%
One finds from this that a field configuration $\Sigma_6 \to \tilde{X}$ on a 6-dimensional worldvolume $\Sigma_6$ is equivalently
a pair consisting of an ordinary Sigma-model field $\phi \colon \Sigma_6 \to X$ together with a $\phi$-twisted
2-form connection on $\Sigma$. These are 
the components of the ``tensor multiplet'' of fields on the M5. 


\section{The equivariant M5-brane cocycle}
Following general results by Hinich, Getzler, Pridham and Lurie \cite{hinich, getzler, pridham, Lurie-dagX}, we may think of $L_\infty$-algebras as being infinitesimal
higher stacks. In the context of the present article, this is discussed  in \cite{dcct}. Therefore, we can think of
the diagram (\ref{The7CocycleOnHomotopyFiber}) as describing at an infinitesimal level a geometric situation.
The geometric picture is that of a (homotopy) principal bundle $P\to B$ over a base $B$, classified by a morphism $B\to BG$ for some gauge group $G$, together with a morphism $\varphi\colon P\to M$ defined over the total space of the bundle and with values in some $G$-space $M$. As soon as  $\varphi$ is equivariant with respect to the (homotopy) gauge group $G$ action over the total space $P$ of the bundle, the morphism $\varphi$ induces by passing to quotients a morphism $\tilde{\varphi}\colon B\to M/G$, whose homotopy class can be interpreted as an element in the twisted cohomology of $B$ with coefficients in $M/G$. Notice how saying that $M/G$ is a quotient of $M$ by a $G$ action is equivalent to say that $M/G$ fits into a homotopy fiber sequence of the form $M\to M/G\to BG$, and that the $G$-equivariance of $\varphi$ is equivalent to saying that $\tilde{\varphi}$ is a morphism over $BG$. See \cite[section 4]{NSSa} for details.

\medskip
Therefore, in the infinitesimal situation of diagram (\ref{The7CocycleOnHomotopyFiber}), we are faced with the problem of deciding whether the cocycle
$h_3 \wedge \mu_4 +  \tfrac{1}{15}\mu_7\colon  \mathfrak{m}2\mathfrak{brane}\to \mathbb{R}[6]$ is equivariant with respect to the natural infinitesimal homotopy action of $\mathbb{R}[2]$ on $\mathfrak{m}2\mathfrak{brane}$ and for a suitable infinitesimal homotopy action of $\mathbb{R}[2]$ on $\mathbb{R}[6]$.
%
%
%
%
%
%
%
As we are going to show, there is indeed such a canonical action. We now identify the desired higher algebra: 
the Lie 7- algebra $\mathfrak{s}^4$ 
is defined by
$
\mathrm{CE}(\mathfrak{s}^4)=\mathbb{R}[g_4,g_7]
$
with $g_k$ in degree $k$ and with the differential defined by
$$
  d g_4 = 0
  \;\;
  \,,
  \;\;
  d g_7 = g_4 \wedge g_4
  \,.
$$
Indeed, this algebra satisfies the required property.
%
Namely, the Lie 7-algebra $\mathfrak{s}^4$ has a natural structure of infinitesimal $\mathbb{R}[2]$-quotient of $\mathbb{R}[6]$, i.e., there exists a natural homotopy fiber sequence of $L_\infty$-algebras
\begin{equation}\label{hopf}
  \raisebox{20pt}{
  \xymatrix{
    \mathbb{R}[6] \ar[r] \ar[d]& \mathfrak{s}^4 \ar[d]^{p}
    \\
    0\ar[r]& \mathbb{R}[3]\;.
  }
  }
\end{equation}
%
This can be seen by noticing that in the dual semifree differential (bi)-graded algebra picture we have the evident  pushout
$$
  \raisebox{20pt}{
  \xymatrix{
    \mathbb{R}[g_4,g_7] \ar[rr]^{{g_4\mapsto 0}\atop{g_7\mapsto g_7}} &&\mathbb{R}[g_7]
    \\
    \mathbb{R}[g_4]\ar[rr]\ar[u]^{g_4\mapsto g_4}& &\mathbb{R}\ar[u]\;.
  }
  }
$$
That this indeed corresponds to a homotopy fiber sequence of $L_\infty$-algebras can be seen by the argument in \cite[theorem 3.1.13]{LocalObservables}.
%
%
Notice how 
we have implicitly used the fact that $\mathbb{R}[3]$ is a delooping of $\mathbb{R}[2]$ and hence serves as the infinitesimal classifying space for infinitesimal $\mathbb{R}[2]$-bundles.
With the geometric picture described at the beginning of this section in mind, we want now to exhibit an $L_\infty$-algebra morphism $\tilde{\varphi}\colon \mathbb{R}^{10,1\vert \mathbf{32}}\to \mathfrak{s}^4$ over $\mathbb{R}[3]$, inducing the 6-cocycle $\varphi=h_3 \wedge \mu_4 +  \tfrac{1}{15}\mu_7$ by passing to the homotopy fibers. In other words, we want to realize a homotopy commutative diagram of $L_\infty$-algebras of the form
$$
  \raisebox{20pt}{
  \xymatrix{
   \mathfrak{m}2\mathfrak{brane}\ar[d]\ar[ddr]\ar[r]^{\varphi} &\mathbb{R}[6]\ar[ddr]\ar[d]\\
   0\ar[ddr] \ar[r]|(.5)\hole& 0\ar[ddr]|(.5)\hole\\
   & \mathbb{R}^{10,1\vert \mathbf{32}}
	\ar[r]^>>>>>{\tilde{\varphi}}
	\ar[d]
	&
	\mathfrak{s}^4
	\ar[d]
	\\
&	\mathbb{R}[3]
	\ar@{=}[r]
	&
	\mathbb{R}[3]\;.
  }
  }
$$
Passing to the dual picture, we see that this is realized by the commutative diagram of differential (bi-)graded semifree commutative algebras
$$
  \raisebox{20pt}{
  \xymatrix{
   \mathbb{R}[g_4,g_7]\ar[ddr]^(.35){\!\!\!\!{\!\!\!g_4\mapsto 0}\atop{g_7\mapsto g_7}}
   \ar[rr]^-{{g_4\mapsto g_4}\atop{g_7\mapsto h_3 \wedge (g_4 + \mu_4)  + \tfrac{1}{15}\mu_7}} &&\mathrm{CE}({\mathbb{R}}_{\mathrm{res}}^{10,1\vert \mathbf{32}})\ar[r]^-{{h_3\mapsto 0}\atop{g_4\mapsto \mu_4}}_{\sim}\ar[ddr]^(.35){\!\!\!{h_3\mapsto h_3}\atop{g_4\mapsto 0}}
   &\mathrm{CE}({\mathbb{R}}^{10,1\vert \mathbf{32}})
   \\
   \mathbb{R}[b_4]\ar[u]^{g_4\mapsto g_4}\ar[ddr]
    \ar@{=}[rr]|(.23)\hole&& \mathbb{R}[b_4]\ar[u]^{g_4\mapsto g_4}\ar[ddr]|(.5)\hole
    \ar@{=}[r]|(.5)\hole &\mathbb{R}[b_4]\ar[u]_{g_4\mapsto\mu_4}
    \\
   & \mathbb{R}[b_7]\ar[rr]^<<<<<<<<<<<<<<<{g_7\mapsto h_3 \wedge \mu_4 +  \tfrac{1}{15}\mu_7}
	&&
	\mathrm{CE}(\mathfrak{m}2\mathfrak{brane})
	\\
&	\mathbb{R}\ar[u]
	\ar@{=}[rr]
	&&
	\mathbb{R}
	\ar[u] \;,
  }
  }
$$
where $\mathrm{CE}({\mathbb{R}}_{\mathrm{res}}^{10,1\vert \mathbf{32}})$ is the semifree algebra obtained by adding to $\mathrm{CE}({\mathbb{R}}^{10,1\vert \mathbf{32}})$ two generators $h_3$ and $g_4$, in degree 3 and 4 respectively, with differential
$$
  d h_3 := g_4 - \mu_4; \qquad dg_4=0
  \,.
$$
This is manifestly homotopy equivalent to $\mathrm{CE}({\mathbb{R}}^{10,1\vert \mathbf{32}})$: the morphism
from $\mathrm{CE}({\mathbb{R}}_{\mathrm{res}}^{10,1\vert \mathbf{32}})$ to $\mathrm{CE}({\mathbb{R}}^{10,1\vert \mathbf{32}})$ given on the additional generators by
$$
  h_3 \mapsto 0\,, \;\; g_4 \mapsto \mu_4
$$
is a homotopy inverse to the inclusion $\mathrm{CE}({\mathbb{R}}^{10,1\vert \mathbf{32}})\hookrightarrow \mathrm{CE}({\mathbb{R}}_{\mathrm{res}}^{10,1\vert \mathbf{32}})$. We can think of the $L_\infty$-algebra ${\mathbb{R}}_{\mathrm{res}}^{10,1\vert \mathbf{32}}$ as a \emph{resolution} of the super-Minkowski space  ${\mathbb{R}}^{10,1\vert \mathbf{32}}$: from the point of view of homotopy theory of $L_\infty$-algebras, ${\mathbb{R}}_{\mathrm{res}}^{10,1\vert \mathbf{32}}$ and ${\mathbb{R}}^{10,1\vert \mathbf{32}}$ are \emph{the same} space. Similarly, the commutative square on the top right of the diagram shows that from the point of view of the homotopy theory of $L_\infty$-algebras, $g_4\mapsto \mu_4$ and $g_4\mapsto g_4$ are the the same 3-cocycle.

\medskip
The only nontrivial check in the above diagram is the compatibility of the morphism $ \mathbb{R}[g_4,g_7]\to \mathrm{CE}({\mathbb{R}}_{\mathrm{res}}^{10,1\vert \mathbf{32}})$ with the differentials on the generator $g_7$ of $ \mathbb{R}[g_4,g_7]$. One has
\[
g_7\stackrel{d}{\mapsto}g_4\wedge g_4 \mapsto g_4\wedge g_4
\]
and
\[
g_7\to h_3 \wedge (g_4 + \mu_4)  +\tfrac{1}{15} \mu_7 \stackrel{d}{\mapsto} (g_4-\mu_4)\wedge (g_4+\mu_4)+ \mu_4 \wedge \mu_4=g_4\wedge g_4.
\]
As mentioned at the beginning of this section, the morphism $g_4\mapsto g_4$; $g_7\mapsto  h_3 \wedge (g_4+ \mu_4) + \tfrac{1}{15}\mu_7$ defines an element in the $\mathbb{R}[2]$-twisted $L_\infty$-algebra cohomology of the super-Minkowski space ${\mathbb{R}}_{\mathrm{res}}^{10,1\vert \mathbf{32}}$, corresponding to the 6-cocycle $h_3 \wedge  \mu_4 + \tfrac{1}{15}\mu_7$ on $\mathfrak{m}2\mathfrak{brane}$.
%
%
%
Upon passing from $L_\infty$-cocycles to their WZW terms, this is
to refine into a statement about twisted differential cohomology, i.e., to a morphism of smooth stacks. This we turn to next.

\section{Integration to differential cohomotopy}
The main result of the previous section can be summarized as follows: the homotopy pullback diagram (\ref{The7CocycleOnHomotopyFiber}) translates to a lift
$$
  \xymatrix{
     \mathbb{R}^{10,1\vert \mathbf{32}}
    \ar[dr]_{\mu_4}
    \ar[rr]^{\tilde{\varphi}}
    && \mathfrak{s}^4 \ar[dl]
    \\
    & \mathbb{R}[3]
  }\,
$$
of $\mu_4$, realizing $\varphi=h_3 \wedge \mu_4 + \tfrac{1}{15} \mu_7$ as a twisted cocycle on $\mathbb{R}^{10,1\vert \mathbf{32}}
$. Reinterpreting this in terms of WZW Lagrangians, this says that the  homotopy pullback diagram (\ref{HigherGaugedTargetSpacetime}) translates to a lift
$$
  \xymatrix@C=0pt{
    X
    \ar[dr]_{\mathbf{L}_{\mathrm{WZW}}^{\mathrm{M2}}}
    \ar[rr]^{\tilde{\mathbf{L}}_{\mathrm{WZW}}^{\mathrm{M5}}}
    && S^4_{\mathrm{conn}} \ar[dl]
    \\
    & \mathbf{B}^3 U(1)_{\mathrm{conn}}
  }\,
$$
of $\mathbf{L}_{\mathrm{WZW}}^{\mathrm{M2}}$, realizing the M5-brane WZW Lagrangian $\mathbf{L}_{\mathrm{WZW}}^{\mathrm{M5}}$ as a morphism from the super-spacetime $X$ to a suitable stack of fields $S^4_{\mathrm{conn}}$. We now exhibit a construction of the stack $S^4_{\mathrm{conn}}$ and provide a geometric interpretation of it, which consequently leads to interpret  the M5-brane WZW Lagrangian as an element in differential cohomotopy.

\medskip
More precisely, what we want to show is that the stack $S^4_\mathrm{conn}$ may be taken to be a differential refinement of the rational homotopy type $\mbox{$\int$}(S^4)$ of the 4-sphere $S^4$ and that the morphism $S^4_\mathrm{conn}\to\mathbf{B}^3 U(1)_{\mathrm{conn}}$ is a differential refinement of the classifying map $S^4\to K(\mathbb{Z};4)$ for the generator of the fourth cohomology group of $S^4$. Following \cite{Henriques, FSS}, recall that the Sullivan construction integrates an $L_\infty$-algebra $\mathfrak{g}$ to an $\infty$-groupoid (a simplicial space which satisfies Kan's condition on horn fillers) $\flat \exp(\mathfrak{g})$ defined by
$$
  \flat \exp(\mathfrak{g})
  :
  [k]
  \mapsto
  \Omega^1_{\mathrm{flat}}(\Delta^k,\mathfrak{g})
  \,,
$$
where
$$
  \Omega^1_{\mathrm{flat}}(X,\mathfrak{g}) := \mathrm{Hom}_{\mathrm{dgAlg}}(\mathrm{CE}(\mathfrak{g}), \Omega^\bullet(M))
$$
is the set of flat $\mathfrak{g}$-valued differential forms on a given manifold $M$. Then one says that the $L_\infty$-algebra $\mathfrak{g}$ is a Sullivan model for the rational homotopy type of a space $X$ if one has an equivalence of simplicial spaces $ \mbox{$\int$}(X)\otimes \mathbb{R} \; \simeq \flat \exp(\mathfrak{g})$. As a remarkable example, the Lie 7-algebra $\mathfrak{s^4}$ is a Sullivan model for the rational homotopy type of the 4-sphere $S^4$, see \cite{Su}. Moreover, the fiber sequence of higher Lie algebras
(\ref{hopf}) is a shadow of the Sullivan model for the rational homotopy type of the Hopf fibration $S^7\to S^4\to BSU(2)$.

\medskip
Refining the Sullivan construction, one then defines the (higher) stack $\flat$-$\mathfrak{g}$-conn of flat $\mathfrak{g}$-connection, as the stack locally given by
\[
\flat\text{-}\mathfrak{g}\text{-conn}\colon (U,[k]) \mapsto \Omega^1_{\mathrm{flat}}(U \times \Delta^k;\mathfrak{g})
\]
on a smooth (super-)manifold $U$ diffeomorphic to $\mathbb{R}^{m|n}$ for some $m|n$.  The stack $\flat\text{-}\mathfrak{g}\text{-conn}$ is to be thought of as resolution of the locally constant stack $\flat\exp(\mathfrak{g})$, directly generalizing the de Rham resolution of the sheaf of constant real valued functions \cite{dcct}.
It has the advantage that the canonical inclusion of $0$-simplices exhibits then a natural morphism of stacks
$$
  \Omega^1_{\mathrm{flat}}(-,\mathfrak{g}) \to \flat\text{-}\mathfrak{g}\text{-conn}
  \,.
$$
This morphism simply says that a globally defined flat $\mathfrak{g}$-valued form on a smooth (super-)manifold $X$ is in particular a flat $\mathfrak{g}$-connection on $X$ whose underlying bundle is in the trivial topological sector.

\medskip
Let us work out in detail the two examples that will be relevant to the construction of $S^4_{\mathrm{conn}}$.
If $\mathfrak{g}=\mathbb{R}[p+1]$, then $\Omega^1_{\mathrm{flat}}(-,\mathbb{R}[p+1])$ is nothing but the sheaf $\Omega^{p+2}_\mathrm{cl}$ of closed $(p+2)$-forms, while by the Poincar\'e lemma flat $\mathbb{R}[p+1]$-connections are equivalently \v{C}ech $(p+2)$-cocycles with coefficients in the sheaf $\flat \underline{\mathbb{R}}$ of locally constant $\mathbb{R}$-valued functions. In other words, we have a natural equivalence of $(p+2)$-stacks $\flat\text{-}\mathbb{R}[p+1]\text{-conn}\cong \mathbf{B}^{p+2}\flat\underline{\mathbb{R}}$ and
the morphism of stacks
$$
\Omega^{p+2}_\mathrm{cl}\to \mathbf{B}^{p+2}\flat\underline{\mathbb{R}}
$$
is induced at the level of chain complexes by the commutative diagram
\[
\xymatrix{
0\ar[r]\ar[d]&0\ar[r]\ar[d]&\dots\ar[r]&0\ar[r]\ar[d]&\Omega^{p+2}_{\mathrm{cl}}\ar[d]\\
 \Omega^0\ar[r]^d&\Omega^1\ar[r]^d& \cdots \ar[r]^d&\ar[r]^d\Omega^{p+1}\ar[r]^d&\Omega^{p+2}_{\mathrm{cl}}\\
 \flat\underline{\mathbb{R}}\ar[r]\ar[u]&0\ar[r]\ar[u]&\dots\ar[r]&0\ar[r]\ar[u]&0\ar[u]
}
\]
Passing to cohomology, this is nothing but the canonical morphism $\Omega^{p+2}_{\mathrm{cl}}(X)\to H^{p+2}_{\mathrm{dR}}(X)\cong H^{p+2}(X;\mathbb{R})$, for any smooth manifold $X$.
%
The inclusion of $\mathbb{Z}$ in $\mathbb{R}$ induces a morphism of sheaves $\mathbb{Z}\to \underline{\mathbb{R}}$ and so a morphism of stacks $\mathbf{B}^{p+2}\mathbb{Z}\to  \mathbf{B}^{p+2}\flat\underline{\mathbb{R}}$. We may therefore pull this morphism back along the morphism $\Omega^{p+2}_\mathrm{cl}\to \mathbf{B}^{p+2}\underline{\mathbb{R}}$. What one gets is the stack $\mathbf{B}^{p+1}U(1)_\mathrm{conn}$, i.e., one has a homotopy pullback diagram
\begin{equation}\label{bun-pullback}
\xymatrix{
\mathbf{B}^{p+1}U(1)_\mathrm{conn}\ar[r]\ar[d]_{F} & \mathbf{B}^{p+2}\mathbb{Z}\ar[d]\\
\Omega^{p+2}_{\mathrm{cl}}\ar[r]&\mathbf{B}^{p+2}\flat\underline{\mathbb{R}}\;,
}
\end{equation}
where the morphism $F$ is the \emph{curvature}. See \cite{FSS} for details on this homotopy pullback description of $\mathbf{B}^{p+1}U(1)_\mathrm{conn}$.
%
%

\medskip
The second example relevant to our discussion is the following.
The sheaf $\Omega^1_{\mathrm{flat}}(-;\mathfrak{s}^4)$ is the sheaf whose sections over a smooth (super-)manifold $M$ are the pairs $(\omega_4,\omega_7)$ where $\omega_4$ is a closed 4-form on $M$ and $\omega_7$ is a 7-form on $M$ such that $d\omega_7=\omega_4\wedge\omega_4$. By naturality of the Sullivan construction, the 3-cocycle $\mathfrak{s^4}\to \mathbb{R}[3]$ induces a homotopy commutative diagram of stacks
\[
\xymatrix{
\Omega^1_{\mathrm{flat}}(-;\mathfrak{s}^4)\ar[r]\ar[d]&\Omega^1_{\mathrm{flat}}(-;\mathbb{R}[4])\ar@{=}[r]&\Omega^4_{\mathrm{cl}}\ar[d]\\
\flat\text{-}\mathfrak{s}^4\text{-conn}\ar[r]& \flat\text{-}\mathbb{R}[3]\text{-conn}\ar[r]^-{\sim} &\mathbf{B}^4\flat\underline{\mathbb{R}}
}
\]
The top horizontal arrow in this diagram is simply the projection $(\omega_4,\omega_7)\mapsto \omega_4$. The kernel of this map consists of closed 7-forms. We therefore have a fiber sequence
\[
\Omega^7_\mathrm{cl}\to \Omega^1_{\mathrm{flat}}(-;\mathfrak{s}^4)\to \Omega^4_\mathrm{cl}
\]
which can be read as a differential forms version of the Hopf fibration $S^7\to S^4\to BSU(2)\stackrel{c_2}{\to} K(\mathbb{Z},4)$.
The bottom horizontal morphism in the above diagram  is the $\mathbb{R}$-localization of the classifying map $S^4\to K(\mathbb{Z};4)$ for the generator of $H^4(S^4;\mathbb{Z})$, i.e., we have a homotopy commutative diagram of smooth stacks
\[
\xymatrix{
\mbox{$\int$}(S^4)\ar[r]\ar[d]_{-\otimes\mathbb{R}} & \mathbf{B}^4\mathbb{Z}\ar[d]^{-\otimes\mathbb{R}}\\
\flat\text{-}\mathfrak{s}^4\text{-conn}\ar[r]&\mathbf{B}^4\flat\underline{\mathbb{R}}\;.
}
\]
%

\medskip
We can now define the stack $S^4_{\mathrm{conn}}$ by analogy with the homotopy pullback definition (\ref{bun-pullback}) of the stack $\mathbf{B}^{p+1}U(1)_\mathrm{conn}$ 
Namely, looking at the homotopy type $\mbox{$\int$}(S^4)$ of $S^4$ as a locally constant geometrically discrete stack, the $\mathbb{R}$-localization map $\mbox{$\int$}(S^4)\to \mbox{$\int$}(S^4)\otimes \mathbb{R}$ is promoted to a morphism of stacks $\mbox{$\int$}(S^4)\to \flat\text{-}\mathfrak{s}^4\text{-conn}$, which we can pull back along $\Omega^1_{\mathrm{flat}}(-;\mathfrak{s}^4)\to \flat\text{-}\mathfrak{s}^4\text{-conn}$. The stack $S^4_{\mathrm{conn}}$ is then defined as this pullback. By the univeral property of the homotopy pullback we therefore get a canonical morphism $S^4_{\mathrm{conn}}\to \mathbf{B}^{3}U(1)_\mathrm{conn}$ fitting into a homotopy commutative diagram of the form
$$
  \raisebox{20pt}{
  \xymatrix{
  S^4_{\mathrm{conn}}\ar[d]\ar[ddr]\ar[r] &\mbox{$\int$}(S^4)\ar[ddr]\ar[d]\\
   \Omega^1_{\mathrm{flat}}(-;\mathfrak{s}^4)\ar[ddr] \ar[r]|(.5)\hole&  \flat\text{-}\mathfrak{s}^4\text{-conn}\ar[ddr]|(.5)\hole\\
   &  \mathbf{B}^{3}U(1)_\mathrm{conn}
	\ar[r]
	\ar[d]
	&
	\mathbf{B}^4\mathbb{Z}
	\ar[d]
	\\
&	\Omega^4_{\mathrm{cl}}	\ar[r]
	&
	\mathbf{B}^4\flat\underline{\mathbb{R}}\;.  }
  }
$$
Since the stack $\mathbf{B}^4\mathbb{Z}$  is the locally constant geometrically discrete stack defined by the homotopy type $K(\mathbb{Z},4)$, the 
 square
\[
\xymatrix{
S^4_{\mathrm{conn}}\ar[r]\ar[d] &\mbox{$\int$}(S^4)\ar[d]\\
\mathbf{B}^{3}U(1)_\mathrm{conn}\ar[r]&\mathbf{B}^4\mathbb{Z}
}
\]
in the above homotopy commutative diagram
precisely exhibits $S^4_{\mathrm{conn}}$ as a differential refinement of the homotopy type of $S^4$ in analogy to how $\mathbf{B}^{3}U(1)_\mathrm{conn}$ is a differential refinement of the homotopy type $K(\mathbb{Z};4)$.


\medskip
Since the theory of higher homotopy groups of a topological space $X$ is essentially the theory of homotopy classes of maps from spheres to $X$, the theory of homotopy classes of maps from $X$ to spheres is also known as \emph{cohomotopy}. With this terminology, an element in the nonabelian cohomology of $X$ with coefficients in (unstabilized) spheres is called a cohomotopy class for $X$, while a map $f\colon X\to S^k$ representing it can be thought of as a cohomotopy cocycle. This way, one thinks of the M5 WZW Lagrangian $\tilde{\mathbf{L}}_{\mathrm{WZW}}^{\mathrm{M5}}$ as a cocycle in ``differential cohomotopy'' for the super-spacetime $X$. This has been suggested before in \cite[section 2.5]{Sati13}.

\medskip
To conclude this section, notice that
a more ``conservative'' choice of integrated coefficients, i.e., of a target stack for the  M5-brane WZW Lagrangian 
to takes values into, is the following.\footnote{Thanks to Thomas Nikolaus for discussion.} The cup product in integral cohomology gives rise to a morphism of stacks
\[
\cup^2\colon\mathbf{B}^3 U(1)\xrightarrow{\mathrm{diag}}\mathbf{B}^3 U(1)\times \mathbf{B}^3 U(1)\xrightarrow{\cup} \mathbf{B}^7 U(1)\,,
\]
and so we have a long fiber sequence
$$
  \raisebox{20pt}{
  \xymatrix{
    \mathbf{B}^6 U(1)\ar[d]
    \ar[r] & \mathbf{B}^6 U(1)/\!/\mathbf{B}^2 U(1)
    \ar[d]\ar[r]&\ast\ar[d]
    \\
  \ast\ar[r]  & \mathbf{B}^3 U(1)
    \ar[r]^-{\cup^2}
    &
    \mathbf{B}^7 U(1)\;,
  }
  }
$$
where $\mathbf{B}^6 U(1)//\mathbf{B}^2 U(1)$ is \emph{by definition} the homotopy fiber of $\cup^2\colon \mathbf{B}^3 U(1)\to \mathbf{B}^7 U(1)$. This construction can be refined by adding connections to the picture simply by replacing the cup product in integral cohomology with the cup product in Deligne cohomology, see \cite{CupProducts}. This way one gets the
long fiber sequence
$$
  \raisebox{20pt}{
  \xymatrix{
    \mathbf{B}^6 U(1)_{\mathrm{conn}}\ar[d]
    \ar[r] & (\mathbf{B}^6 U(1)/\!/\mathbf{B}^2 U(1))_{\mathrm{conn}}
    \ar[d]\ar[r]&\ast\ar[d]
    \\
    \ast\ar[r]& \mathbf{B}^3 U(1)_{\mathrm{conn}}
    \ar[r]^-{\cup^2}
    &
    \mathbf{B}(\mathbf{B}^6 U(1)_{\mathrm{conn}})\;,
  }
  }
$$
where $(\mathbf{B}^6 U(1)/\mathbf{B}^2 U(1))_{\mathrm{conn}}$ is defined by the homotopy fiber poduct on the right.
On curvature forms this homotopy fiber product imposes again the condition $d G_7 = G_4\wedge G_4$, so we have found a geometric way of imposing this constraint that does not involve the more sophisticated stack $S^4_{\mathrm{conn}}$. One could then define the M5-brane WZW term to be a
lift
$$
  \raisebox{20pt}{
  \xymatrix@C=0pt{
    X
    \ar[dr]_{\mathbf{L}_{\mathrm{WZW}}^{\mathrm{M2}}}
    \ar[rr]^-{\widehat{\mathbf{L}}_{\mathrm{WZW}}^{\mathrm{M5}}}
    && (\mathbf{B}^6 U(1)/\mathbf{B}^2 U(1))_{\mathrm{conn}} \ar[dl]
    \\
    & \mathbf{B}^3 U(1)_{\mathrm{conn}}
  }}\,
$$
of $\mathbf{L}_{\mathrm{WZW}}^{\mathrm{M2}}$.
These would actually be a strictly less general solution than the one provided by the stack $S^4_{\mathrm{conn}}$. Namely, for dimensional reasons the image of the canonical map
$S^4_{\mathrm{conn}} \to \mathbf{B}^2 U(1)_{\mathrm{conn}}$ under $\cup^2$ has a trivializing homotopy,
and so the universal property of the homotopy fiber induces a factorization of $\widehat{\mathbf{L}}_{\mathrm{WZW}}^{\mathrm{M5}}$ through $\tilde{\mathbf{L}}_{\mathrm{WZW}}^{\mathrm{M5}}$:
$$
  \raisebox{20pt}{
  \xymatrix{
    X\ar[rr]^{\tilde{\mathbf{L}}_{\mathrm{WZW}}^{\mathrm{M5}}}\ar[drr]\ar@/^3pc/[rrrr]^{\widehat{\mathbf{L}}_{\mathrm{WZW}}^{\mathrm{M5}}}
    &&S^4_{\mathrm{conn}}
    \ar[rr]
    \ar[d]
    &&
    (\mathbf{B}^6 U(1)/\mathbf{B}^2 U(1))_{\mathrm{conn}}\;.
    \ar[dll]
    \\
    && \mathbf{B}^3 U(1)_{\mathrm{conn}}
  }
  }
$$

\section{Extension to frame bundles}

So far we considered the cocycles $\mu_{p+2}$ and the  as cocycles on the (extended) super-Minkowski spacetime.
In fact, since the cocycles $\mu_{p+2}$ are
Lorentz-invariant they extend to cocycles on the super-Poincar{\'e} (super-)Lie algebra
along the inclusion
$$
  \mathbb{R}^{10,1\vert \mathbf{32}}
  \hookrightarrow
  \mathfrak{iso}(\mathbb{R}^{10,1\vert \mathbf{32}})
  \,.
$$
In terms of semifree algebras, the super-Poincar{\'e} Lie algebra is obtained by adding to $\mathrm{CE}(\mathbb{R}^{10,1\vert \mathbf{32}})$ the
additional `rotational' generators $\{\omega^{a}{}_b\}$
corresponding to the Lorentz Lie algebra, with differentials
$$
  d \psi = \tfrac{1}{4}\omega_{a b} \wedge\Gamma^{a b} \psi
  \;\;
  \,,
  \;\;
  d e^a = \omega^a{}_b \wedge e^b + \overline{\psi}\wedge \Gamma^a \psi
  \;\;
  \,,
  \;\;
d\omega^{a}{}_b= \omega^{a}{}_c \wedge \omega^c{}_b\;.
$$ 
The inclusion $ \mathbb{R}^{10,1\vert \mathbf{32}}
  \hookrightarrow
  \mathfrak{iso}(\mathbb{R}^{10,1\vert \mathbf{32}})$ is given, in the dual semifree algebras picture, by mapping to zero the added generators $\omega^{a}{}_b$.
Therefore, we see that there are possibly more general cocycles on the super-Poincar{\'e} Lie algebra which restrict to a given cocycle on
on super-Minkowski space. In particular, we see that if $\lambda_3(\omega)$ and $\lambda_7(\omega)$ are a Lorentz 3-cocycle and a Lorentz 7-cocycle (i.e., closed polynomials of degree 3 and 7 in the variables $\omega^{a}{}_b$ , respectively), then $(h_3 + \omega^{\wedge 3})\wedge (g_4+ \mu_4) +  \tfrac{1}{15}\mu_7 + \omega^{\wedge 7}$ is a $\mathfrak{s}^4$-valued 7-cocycle on the super-Poincar{\'e} Lie algebra inducing the 7-cocycle $h_3 \wedge (g_4+ \mu_4) + \tfrac{1}{15}\mu_7$ on $\mathbb{R}^{10,1\vert \mathbf{32}}$.
Up to normalization, the only such Lorentz cocycles are the traces $\lambda_3(\omega)=\mathrm{tr}( \omega^{\wedge 3})$ and $\lambda_7(\omega)=\mathrm{tr}( \omega^{\wedge 7})$
of the third and the seventh wedge power of the matrix-valued form $\omega$, respectively. This way we obtain the 2-parameter family of 7-cocycles
$$
 \mathfrak{iso}(\mathbb{R}^{10,1\vert \mathbf{32}})\xrightarrow{(h_3 + \alpha\mathrm{tr}( \omega^{\wedge 3}))\wedge (g_4+ \mu_4) +  \tfrac{1}{15}\mu_7 +\beta \mathrm{tr}( \omega^{\wedge 7})} \mathfrak{s}^4
$$
lifting $h_3 \wedge (g_4+ \mu_4) + \tfrac{1}{15}\mu_7$.

\medskip
Due to the isomorphism
$\mathfrak{iso}(\mathbb{R}^{10,1\vert \mathbf{32}}) \simeq \mathbb{R}^{10,1\vert \mathbf{32}}\ltimes \mathfrak{so}(10,1)$,
a natural choice of globalization of these cocycles is over the (Spin-)frame bundle $\mathrm{Fr}(X) \to X$ of super-spacetime
$X$, where $\mathrm{tr}( \omega^{\wedge 3})$ is globalized as
a parameterized WZW term over the $\mathrm{Spin}$-fibers of the frame bundle. According to \cite{WZWterms}
this is a parameterized WZW term as considered in \cite{DistlerSharpe}, whose existence trvializes the
class $\tfrac{1}{2}p_1$ of the frame bundle. Once such a trivialization is given, $\mathrm{tr}( \omega^{\wedge 7})$ globalizes 
to a parametrized degree-7 WZW term 
related to the calss $p_2$. This 
can be viewed as a Fivebrane  version \cite{SSS2} of the String setting discussed in \cite{DistlerSharpe}.


\end{document}